\documentclass{ws-ijmpd}
\usepackage[super,compress]{cite}
\begin{document}

\markboth{H. Shinkai, M. Takamoto, \& H. Katori}
{Transportable Optical Lattice Clocks and General Relativity}

\title{%
Transportable optical lattice clocks and general relativity
%TRANSPORTABLE OPTICAL LATTICE CLOCKS \\AND GENERAL RELATIVITY
\footnote{This paper is to be published in International Journal of Modern Physics D (2025), and 
also in the book {\em  ``One Hundred and Ten Years of General Relativity: From Genesis and Empirical Foundations to Gravitational Waves, Cosmology and Quantum Gravity}," edited by Wei-Tou Ni (World Scientific, Singapore, 2025).}
}

\author{HISAAKI SHINKAI}
\address{Faculty of Information Science and Technology, Osaka Institute of Technology\\
Kitayama 1-79-1, Hirakata City, Osaka Pref., 573-0196, Japan\\
hisaaki.shinkai@oit.ac.jp}

\author{MASAO TAKAMOTO}
\address{Quantum Metrology Laboratory, RIKEN, \\
Hirosawa 2-1, Wako-shi, Saitama Pref., 351-0198, Japan\\
takamoto@riken.jp}

\author{HIDETOSHI KATORI}
\address{Department of Applied Physics, Graduate School of Engineering, The University of Tokyo, \\
Yayoi 2-11-16, Bunkyo-ku, Tokyo, 113-8656, Japan\\
katori@amo.t.u-tokyo.ac.jp}
\address{Quantum Metrology Laboratory, RIKEN, \\
Hirosawa 2-1, Wako-shi, Saitama Pref., 351-0198, Japan\\
hkatori@riken.jp}

\maketitle

%draft \today \\

%\begin{history}
%\received{Day Month Year}
%\revised{Day Month Year}
%\end{history}

\begin{abstract}

Optical lattice clocks (OLCs) enable us to measure time and frequency with a fractional uncertainty at $10^{-18}$ level, which is 2 orders of magnitude better than Cs clocks. In this article, after briefly reviewing OLCs and the history of testing the fundamental principles of general relativity, we report our experiments of measuring the gravitational redshift between RIKEN and The University of Tokyo, and at Tokyo Skytree using transportable OLCs. We also discuss a couple of future applications of OLCs, such as detecting gravitational waves in space and relativistic geodesy.  The possibility of testing second-order parametrized post-Newtonian potential around the Earth is also mentioned.
\end{abstract}


%\ccode{PACS numbers:}

%\tableofcontents
%=================================================
\section{Introduction}
%=================================================

When Einstein obtained the final form of General Relativity (GR) in 1915, there was only one evidence to believe this theory,
that was the perihelion movement of Mercury, which made Einstein confirm the correctness of the theory.  The measurement of bending light near the Sun at the total eclipse in 1919 made people believe the theory.  
However, the physical phenomena described by GR are too far from everyday life so that the theory was mainly studied from mathematical viewpoints for more than 50 years.
GR predicts many new phenomena, such as black-holes, expanding Universe, and gravitational wave; those are the current main topics in astrophysics, while Einstein is known to have struggled with these three predictions since these are out of his imagination as real physics. 

GR is, mathematically, the simplest theory to express gravity as the result of curved space-time. Amazingly, this theory has beaten all the alternatives over a century.  However, we believe the theory of gravity and quantum should be integrated at some level in order to explain all the nature.  Along this direction, we know the Einstein equation of general relativity is not satisfactory, so that many attempts to construct `quantum theory of gravity' are underway.  The test of GR is, therefore, quite important for developing physics to the next stage. 

The reason why GR is now being discussed in various areas of science is largely due to technological advances that have made it possible to extract information from distant Universe and to conduct experimental tests on the Earth. One of such technology is the development of precise clocks.  In this article, we introduce the optical lattice clock (OLC), one of the current state-of-the-art clocks proposed by one of the authors in 2001\cite{Katori2001}, which has two-order of magnitude more precise than Cs atomic clocks, and discuss how this clock contributes to the science of GR.  

Atomic clocks steer the frequency of local oscillators, such as cavity-stabilized lasers, by referencing atomic transitions.  The stability of such atomic clocks is limited by the quantum projection noise given by the number of atoms. 
By interrogating thousands of atoms trapped at the anti-nodes of a standing-wave laser, which is referred to as an optical lattice, and by eliminating the a.c. Stark shift perturbation by tuning the trapping laser to the magic frequency \cite{Katori2001,Katori2003,Takamoto2003}, the OLCs achieve \cite{Ushijima2015} high stability and precision approaching $10^{-18}$.  By applying an operational magic condition, the clock uncertainty of $10^{-19}$ is in scope \cite{Katori2015}.
The general introduction of OLC is given in \S \ref{sec:OLC}.

According to special relativity, a moving clock ticks slower for an observer at rest.  The ratio is $1/\sqrt{1-(v/c)^2}$, where $v$ and $c$ are relative speed and the speed of light, respectively. 
%The frequency $\nu$ at the rest frame will be $\nu'$ at the moving frame with velocity $v$ as $\nu'=\nu (1-v/c)/\sqrt{1-(v/c)^2}$. 
For $v=10~$m/s, this fractional change of clocks $\Delta \nu / \nu $ is about $5.5\times 10^{-16}$ (see Table.\ref{table1} also), and this was demonstrated using Al$^+$ clocks by Chou {\it et al.} \cite{Chou2010} for $v $ over a couple of m/s. 
According to GR, a clock ticks slower for an observer at deepergravitational potential.  At the lowest order, 
the frequency difference between two clocks which are located at the potential difference $\Delta U$ is given by $\Delta \nu / \nu = \Delta U/c^2$. 
On the surface of the Earth,  this fractional change is about $1.1\times 10^{-16}$ for an altitude difference of $\Delta h = 1$~m.  Therefore OLC can measure the difference of time ticks at a centimeter-scale altitude, and it allows testing Einstein's equivalence principle (EEP), which is the starting principle of GR (\S \ref{sec:EP}).

\begin{table}[htb]
\tbl{Measurable relative speed and height (at the ground of the Earth) with clocks with uncertainty $\Delta \nu/\nu$. }
{\begin{tabular}{@{}ccc@{}} %\toprule
\hline
$\Delta \nu/\nu$ & measurable speed (m/s) & measurable height (m) \\
\hline 
$10^{-15}$ & 13.4~~ & 9.18~~~~ \\
$10^{-16}$ & ~4.24~ & 0.918~~~ \\
$10^{-17}$ & ~1.34~ & 0.0918~~ \\
$10^{-18}$ & ~0.424 & 0.00918~ \\
$10^{-19}$ & ~0.134 & 0.000918 \\
\hline
\end{tabular}\label{table1}}
\end{table}

In 2016, Takano et al.\cite{Takano2016} reported a comparison of OLCs at RIKEN and The University of Tokyo where these two have $\Delta h = 15$~m difference in altitude, which made a constraint to the violation of Local Position Invariance (LPI), one of the main idea of EEP.  In 2020, we  \cite{Takamoto2020} reported a comparison at the broadcasting tower, Tokyo Skytree, where two transportable OLCs placed with $\Delta h = 450$~m difference in altitude, which made a two-order stronger test to LPI ever at the ground level.  These experiments are summarized in \S \ref{sec:SkyTree}. 

The technology of OLC will open new phase of our life and make experiments available for fundamental physics. We discuss some topics in \S \ref{sec:Future}.
Real-time and precise geopotential measurements at the centimeter level will open up new applications in future geopotentiometry\cite{Denker2017}, including seismology and volcanology \cite{Tanaka2021}.
If we place OLCs in space around the Earth, then we will have benefits for long-term stability of Global Navigation Satellite System (GNSS). We also mention the possibility for testing parametrized post-Newtonian (PPN) potential at the second-order around the Earth. 
Finally, we introduce our proposal of new gravitational wave observatory, which we coined {\it Interplanetary Network of Optical Lattice Clocks (INO)}\cite{Ebisuzaki2019}.

%=================================================
\section{Transportable optical lattice clocks \label{sec:OLC}}
%=================================================

This section introduces the principle of OLCs and describes the technological development of transportable OLCs for practical applications.

\subsection{Device and technologies}
\begin{figure}[tbp]
%\vspace*{-2.1cm}
\begin{center}
\hspace*{0.6cm}
\includegraphics[width=\linewidth]{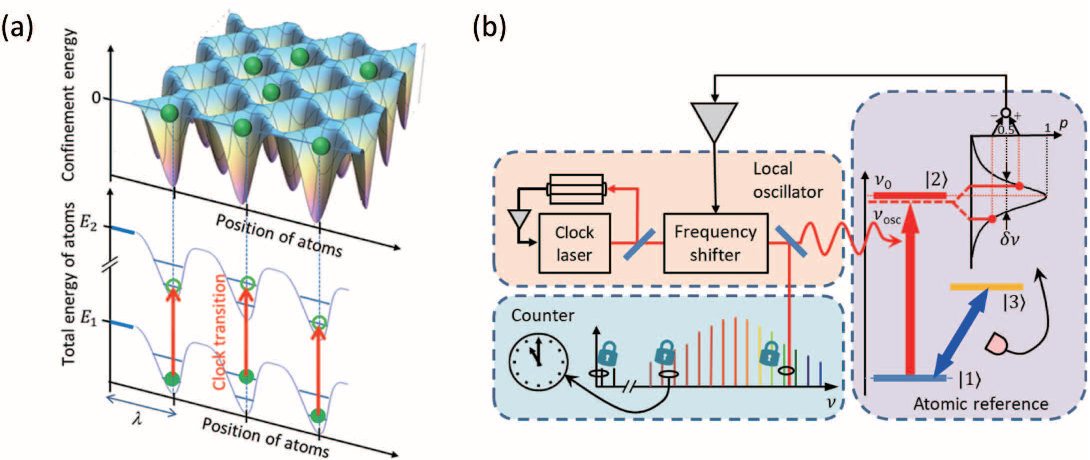}
\end{center}
%\vspace*{2cm}
\caption{A schematic of an optical lattice clock.
(a) Optical lattice potential given by the ac Stark shift is canceled on the clock transition. 
(b) A schematic of an optical atomic clock. A narrow-linewidth laser, stabilized to an optical reference cavity, excites the clock transition $|1\rangle \rightarrow |2\rangle$ of atoms. The excitation probability $p$ is measured by observing the fluorescence on the electric-dipole allowed transition $|1\rangle \rightarrow |3\rangle$, which is used to feedback-control the frequency shifter to keep the clock laser frequency $\nu_{\mathrm{osc}}$ resonant with the clock transition frequency $\nu_0$.  The optical frequency may be down-converted to an RF using a frequency comb.
\label{fig_OLC}
}
\end{figure}

An OLC is an atomic clock based on neutral atoms trapped in a standing wave of lasers, as shown in Fig.~\ref{fig_OLC}(a). 
Subwavelength confinement of atoms by the lattice (Lamb-Dicke regime\cite{Dicke1953})  enables high-precision spectroscopy unaffected by the Doppler and the photon recoil shift. 
Although the a.c. Stark effect caused by the electric field of the lattice laser shifts the frequency of the clock transition, the shift can be suppressed by tuning the lattice laser to a frequency called the ``magic frequency'' at the lowest order \cite{Katori2001}. 
%Although the ac Srark shift of the lattice laser causes a frequency shift on the clock transition, the shift can be suppressed by tuning the lattice laser to a frequency called the ``magic frequency'' at the lowest order \cite{Katori2001}. 
Furthermore, by tuning the frequency and intensity of the lattice laser to an optimum condition called the ``operational magic condition'', the a.c. Stark effect can be compensated to higher orders and can be reduced to $10^{-19}$ \cite{Katori2015,Ushijima2018,McGrew2018}.

Figure~\ref{fig_OLC}(b) illustrates a schematic of the operation of an optical lattice clock.
A narrow-linewidth laser, prestabilized to an optical reference cavity, excites the clock transition $|1\rangle \rightarrow |2\rangle$ of atoms. 
The excitation probability $p$ is determined by observing the fluorescence on the electric-dipole allowed transition $|1\rangle \rightarrow |3\rangle$ and is used to feedback-control the frequency shifter to keep the clock laser frequency $\nu_{\mathrm{osc}}$ resonant with the clock transition frequency $\nu_0$. 
The laser stabilized to the clock transition is down-converted to an RF signal accessible to electronics by an optical frequency comb \cite{Hansch2006, Hall2006}, generating the ``optical second'' based on the optical frequency of atomic transition.

The excitation probability $p$ includes the uncertainty introduced by the projection measurement of the quantum state after clock excitation, called quantum projection noise (QPN)\cite{Itano1993}, which sets the standard quantum limit of instability of atomic clock as given by 
$\sigma_y (\tau) = \delta \nu / (2 \kappa \nu_0) \sqrt{T_c/(N_c \tau)}$, 
where $\tau$ is the averaging time, $\delta \nu$ the linewidth, $\nu_0$ the transition frequency, $N_c$ the number of atoms interrogated per cycle, $\kappa$ coefficient factor of order unity, $T_c$ the cycle time. 
To improve the QPN limited instability, observing atomic spectra having a high-quality factor $Q=\nu_0/\delta \nu$ for many atoms $N_c$ is crucial.
In state-of-the-art optical clocks, the quality factor amounts to $Q \sim 10^{15}$ by employing the clock transition with a natural linewidth of much less than a Hz, practically limited by a laser linewidth of sub-Hz. 
The OLC allows such high-$Q$ transitions to be interrogated with many atoms in the lattice, drastically reducing quantum projection noise and achieving high stability.

%On the other hand, it actually limits the stability limit due to the stability of the pumping laser, called the Dick effect [Dick].
%The stability in recent years is ranked.
%Whereas it takes about a week to measure 18 digits with a single ion clock, it can be done in one hour with an optical lattice clock. 
%Thus, it can be used not only as a time-frequency standard, but also as a measurement tool that can make useful measurements for a wider range of more practical measurements. 

%\clearpage

Since the proposal of the scheme in 2001 \cite{Katori2001}, remarkable progress of research improved the fractional uncertainties of the clocks to $10^{-18}$ or below \cite{Ushijima2015,McGrew2018,Aeppli2024}, which is two orders of magnitude better than those of cesium atomic clocks that currently defines the second in SI units (International System of Units). 
Such significant improvement in OLCs motivates the redefinition of the second by the ``optical second'', to be scheduled for 2030 \cite{Dimarq2024}.

\subsection{Transportable clocks}\label{subsec_TOC}
Once a high-precision clock is established in a laboratory environment, the next challenge is to make it more compact and robust to operate outside the laboratory for practical applications. 
We have developed a transportable OLC using strontium (${}^{87}$Sr) atoms.
The scheme of OLC applies to alkaline earth metal (like) atoms with two electrons in the outermost shell.
The transitions of Sr atoms relevant to clock operation are in the visible region, where semiconductor lasers are available, allowing the development of a compact and robust laser system.
Thus, the Sr atom is a good candidate for transportable clocks for field applications.

\begin{figure}[tpb]
%\vspace*{-2cm}
\begin{center}
\hspace*{1cm}
\includegraphics[width=\linewidth]{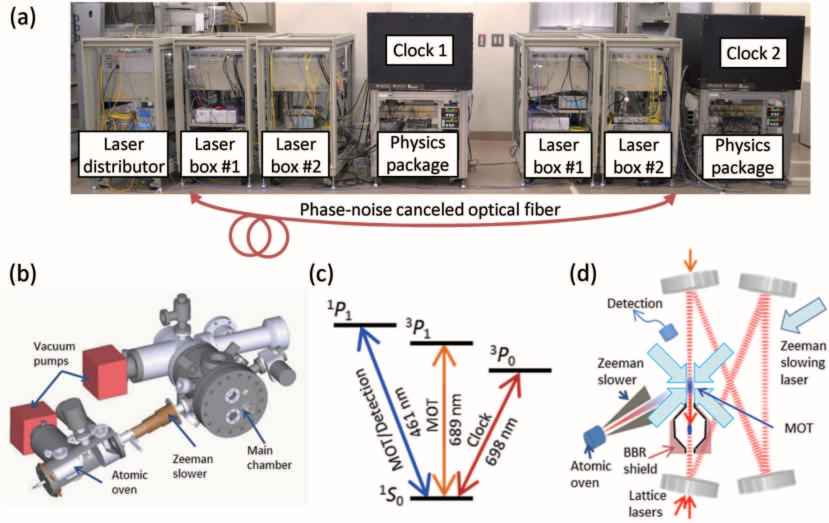}
\end{center}
%\vspace*{2.3cm}
\caption{Transportable OLCs using ${}^{87}$Sr atoms.
(a) Two transportable OLCs. The frequencies of the clocks are compared via a phase-noise-canceled optical fiber.
(b) A schematic of the vacuum chamber inside the physics package.
(c) The relevant energy diagram of Sr atoms.
(d) A schematic of the clock spectroscopy in a bow-tie optical cavity for the lattice.
\label{fig_TOC}
}
\end{figure}

The transportable clock consists of a physics package and two laser boxes, as shown in Fig. \ref{fig_TOC}(a).
The details of the system and its operation are reported in refs. \cite{Takamoto2020, Ohmae2021} .
The lasers for cooling and trapping, lattice, and spectroscopy are supplied from laser boxes to a physics package via optical fibers. 
Two clocks can be connected by a phase-noise-canceled optical fiber controlled by a distributor box to compare their frequencies.

The physics package includes an ultra-high vacuum chamber in which the clock spectroscopy is performed (Fig.~\ref{fig_TOC}(b)). 
Sr atoms from a heated oven are decelerated by a counter-propagating Zeeman slowing laser.
After deceleration, atoms arriving at the center of the main chamber are trapped and cooled down to a few mK by a magneto-optical trap (MOT) on the dipole-allowed $\mathrm{5s^2~{}^1S_0 - 5s5p~{}^1P_1}$ transition (Fig.~\ref{fig_TOC}(c)).
The atoms are further cooled down to a few $\mu \mathrm{K}$ by a narrowline MOT on the spin-forbidden $\mathrm{5s^2~{}^1S_0 - 5s5p~{}^3P_1}$ transition, and loaded into a lattice with a depth of $\sim 20~\mu \mathrm{K}$.

To operate Sr-based OLCs at $10^{-18}$ uncertainty, reduction of the blackbody radiation (BBR) shifts \cite{Nicholson2015,Ushijima2015,McGrew2018} and the higher-order light shifts \cite{Katori2015,Ushijima2018,McGrew2018} is of prime concern. 
The BBR shift is the a.c. Stark effect caused by the BBR emitted from surrounding walls. 
At room temperature of 300~K, the BBR shift amounts to $-5.3 \times 10^{-15}$ for Sr atom.
Applying a small-sized BBR shield as depicted in Fig.~\ref{fig_TOC}(d), the environmental temperature in the spectroscopy region is controlled at 245~K by a four-stage Peltier cooler to reduce the uncertainty of the shift to $3 \times 10^{-18}$. 
In addition, we reduce the a.c. Stark shift of lattice laser to $1 \times 10^{-18}$ by tuning the lattice laser frequency and intensity to the operational magic condition, with polarization parallel to the bias magnetic field \cite{Ushijima2018}.
These parameters compensate the multipolar- and hyperpolarizability-induced a.c. Stark shift by the electric-dipole a.c. Stark shift \cite{Katori2015}. 
To accommodate both the well-defined environmental temperature and lattice intensity, we install a bow-tie cavity for the lattice laser, where detuning of the counter-propagating-laser frequencies allows transporting atoms into the BBR shield.

The clock transition $\mathrm{5s^2~{}^1S_0 - 5s5p~{}^3P_0}$ of atoms inside the BBR shield is interrogated by the clock laser introduced along the lattice axis.
The atoms are transported back to the MOT position and the excitation probability is determined from the fluorescence on the $\mathrm{5s^2~{}^1S_0 - 5s5p~{}^1P_1}$ transition.
According to the excitation probability, the frequency of the clock laser is stabilized to the resonance of the clock transition.
We developed two such clocks and compared their frequencies to confirm the agreement of the clocks with $5\times 10^{-18}$ uncertainty in the laboratory environment.
Such transportable clocks allow conducting experiments outside the laboratory as described in \S \ref{subsec_Skytree}.

%=================================================
\section{Equivalence Principle and General Relativity \label{sec:EP}}
%=================================================
In \S \ref{sec:SkyTree}, we introduce the experiments using OLCs to measure gravitational redshift (\S \ref{sec_gravredshift}), a part of testing the equivalence principle.  
We review the background idea \cite{Will1993Book,Will2014LRR,Will2010AmJP} and past experiments in this section. 
%--------------------------------------------------------------------------------------
\subsection{Equivalence principles} 
%--------------------------------------------------------------------------------------
The starting points of GR is the {\it equivalence principle} (EP).  If this principle is violated, then the usage of tensors and/or the description of space-time using Riemannian geometry will lose its basics. 

There are three expressions of equivalence principle \cite{Will1993Book,Will2014LRR}; weak EP (WEP), Einstein's EP (EEP), and strong EP (SEP).  The statements can be summarized as follows. 
\begin{itemize}
\item {Weak equivalence principle} (WEP) consists of two conditions: 
\begin{itemize}
\item[(i)] {\it ``universality of free fall'' }: 
A sufficiently small object falls with the same acceleration in a gravitational field, regardless of its composition or mass. 
\item[(ii)] {\it ``equivalence of inertial mass and gravitational mass"}: 
The ratio of inertial mass to gravitational mass is constant for all objects.
\end{itemize}
\item {Einstein's equivalence principle} (EEP) consists of two conditions: 
\begin{itemize}
\item[(i)] WEP is satisfied.
\item[(ii)] In a local experiment where gravity does {\it not} act, the following two conditions are satisfied. 
\begin{itemize}
\item {\it Local Lorentz invariance} (LLI): 
Experimental results do not depend on the speed of the system where the experiment is performed. 
\item 
{\it Local position invariance} (LPI): 
Experimental results do not depend on the location or time of the experiment.
\end{itemize}
\end{itemize}
\item {Strong equivalence principle} (SEP) consists of two conditions: 
\begin{itemize}
\item[(i)] WEP is satisfied.
\item[(ii)] Any local experimental results coupled {\it with gravity} satisfy LLI and LPI.
\end{itemize}
\end{itemize}
Verification experiments have been conducted for each, but we should mention that these are only for verifying equivalence principle, and not for whole the theory of relativity.
%--------------------------------------------------------------------------------------
\subsection{Test of equivalence principles \label{sec:EPtest}} 
%--------------------------------------------------------------------------------------
\paragraph{Test of WEP} ~\\
Direct test of WEP can be achieved by free-falling two objects with different compositions. If the WEP is violated, there will be a difference in the resulting acceleration. In other words, the result shows whether inertial mass, $m_{\rm I}$, and gravitational mass, $m_{\rm G}$,  are equivalent or not. 
%$m_{\mbox{\footnotesize inertia}} \neq whether m_{\mbox{\footnotesize gravity}}$
The standard test is to compare the free fall accelerations of each two objects, $a_1, a_2$ and calculate the ratio, 
\begin{equation}
\eta_E\equiv 2\frac{|a_1-a_2|}{a_1+a_2}=
\left(\frac{m_{\rm G}}{m_{\rm I}}\right)_1-\left(\frac{m_{\rm G}}{m_{\rm I}}\right)_2=\Delta \left(\frac{m_{\rm G}}{m_{\rm I}}\right)
%, \qquad \mbox{where~~} a_X=\frac{m_{\rm G}}{m_{\rm I}}} \mbox{~~for~mass~$X=1,\,2$}.
\label{eq_eta}
\end{equation}
This is the ratio $\eta_E$ that E\"otv\"os defined. If $\eta_E=0$, the quantities are equivalent.
\begin{itemize}
\item G. Galilei confirmed this with an uncertainty of $10^{-3}$ in experiments using free-falling balls, balls rolling down slopes, and a pendulum.
\item E\"otv\"os confirmed it with an accuracy of $10^{-9}$ using a torsion balance. Experiments using torsion balances have been applied to experiments using the sun as a gravity source, and in 2012
%\footnote{Williams, J. G., Turyshev, S. G., \& Boggs, D. H. 2012, Classical and Quantum Gravity, 29, 184004\\Murphy, T. W., Jr., Adelberger, E. G., Battat, J. B. R., et al. 2012, Classical and Quantum Gravity, 29,184005}
it reached an uncertainty of $10^{-13}$ (ref. \cite{Wagner2012}), while this accuracy is said to be the limit in ground experiments \cite{1501.01644}.
\item WEP was also tested by using an atomic interferometer (a matter wave interferometer using the ${}^{87}$Rb--${}^{85}$Rb transition). For example, even if the vibrational acceleration of the device is about $1\times 10^{-3}g$ ($g$ is the gravitational acceleration),
%Common-mode vibration noise rejection system allows
Test of WEP  shows that the difference between isotopes was limited to about $\Delta g/g=(1.2\pm 3.2)\times 10^{-7}$ (ref.\cite{Bonnin2013}).
\item The current most vigorous test is the MICROSCOPE mission, which measures the difference in the accelerations of test masses of different compositions in a drag-free satellite around the Earth.  The mission concluded that there is no violation of WEP at the level of $\eta_E \sim 10^{-15}$ (ref. \cite{Microscope2022}).
\end{itemize}

Most of the models that explain the accelerated expansion of the Universe and dark matter by
modifying the theory of gravity require a certain 
minimum value of $\eta_E\approx10^{-18}$.
 Threfore 
%naturally constructed with $\eta \leq 10^{-13}$, but in order to solve the problem,  so that it is necessary that $10^{-18} \leq \eta_E$, and 
it is desirable to verify $\eta_E$ to this level if these approaches are right \cite{Damour2012, STEP}.

\paragraph{Test of EEP} ~\\
WEP yields the result that motion due to gravity and inertia cannot be distinguished, but Einstein reversed this logic and stated, ``a system under a uniform gravitational field can be regarded as a system in motion with uniform acceleration,'' which he used as the guiding principle when constructing GR. 
If we take this as a starting point, the uniformity of free fall becomes a natural result, and the free-falling coordinate system can be regarded as an inertial frame ({local inertial frame}).

EEP postulates the following conditions for gravitational theory: 
\begin{itemize}
\item[(1)] Spacetime is described by a symmetric metric tensor. (We can set the affine connection zero locally.) 
% (= affine can be set to 0)
\item[(2)] The trajectory of a freely falling test particle becomes the geodesic curve of its metric.
% (=uniform linear motion in a free-falling system)
\item[(3)] In a free-falling system, physical laws other than gravity are described by special relativity.
\end{itemize}
%\item
Physical phenomena that result from EEP include the bending of light in a gravitational field and the 
frequency shift of light in a gravitational field.  The latter is our target and is explained in the next. 

The test of LLI, {\it i.e.} the test of no velocity dependency, is equivalent to the test of special relativity. The standard way is to show there is no special direction in the Universe, using a parameter $\delta$ which states the difference between the effective speed of light, $c$, and the possible maximum speed of a particle, $c_0$, 
\begin{equation}
\delta =\left|1 -\frac{c^2_0}{c^2} \right| .  \label{eq_delta}
\end{equation}
If $\delta=0$, then LLI is satisfied. 

The initial interferometer experiment by Michelson and Morley can be regarded as the test of LLI at the level of $\delta < 10^{-3}$. 
A measurement of a time-dependent quadrupole splitting of Zeeman levels (due to the violation of LLI; an anisotropy of inertial mass) with Earth's sidereal frequency shows $ \delta < 3 \times 10^{-21}$ (ref. \cite{Chupp1989}).  
LLI tests using atomic physics were reviewed in ref. \cite{Safronova2018}.
%宇宙線のGZK cutoff 以上の領域で，宇宙マイクロ波背景放射に対する地球の運動(370~km/s)を用いる．($\delta< 1.7 \times 10^{-25}$)

Test of LPI shows that the same results can be obtained even if the experiment is performed at different time or locations.
GR predicts the dilation of time in a deeper gravitational potential; this is referred to as gravitational redshift.
Measuring the gravitational redshift corresponds to observing the time scales at different inertial frames, which can be regarded as a test of LPI. 

LPI says that the gravitational redshift between two clocks (located at positions 1 and 2) 
depends only on the change of the potential, {\it i.e.} the universality of gravitational redshift. 
Using the clock frequency $\nu$ and its difference, $\Delta \nu = \nu_2-\nu_1$, it can be written as 
$\Delta \nu /\nu_1 =  \Delta U /c^2$.  
The parameter used for LPI test is $\alpha$, which is introduced as 
\begin{equation}
\frac{\Delta \nu}{\nu_1}=(1+\alpha) \frac{\Delta U}{c^2}. \label{eq_alpha}
\end{equation}
The parameter $\alpha$ states the difference from GR (If GR,  $\alpha=0$).  See  \S \ref{sec_gravredshift} for more. 

%Furthermore, since LPI asserts that physical laws have no time dependence, the investigation of long-term fluctuations in fundamental physical constants can also be said to be a verification of LPI. The time rate of change of the fine structure constant $\alpha$ is reported to be less than $1.3 \times 10^{-16}$~/yr\cite{Leefer2013} from a long-term comparison of atomic clocks (\ (See also Table 1 in cite{Will2014LRR}).

\paragraph{Test of SEP} ~\\
While the statement of EEP mentions ``local experiments that gravity does not affect'' and excludes those with gravity and/or self-gravity, SEP generalizes the statement ``for any experiment''. 

If SEP holds, then the conclusion is that ``the gravitational theory must be described only in terms of metrics.'' 
Modification of the theory of gravity by higher-order curvature or higher dimensions satisfies SEP, while the theory with scalar and/or vector field does not. 
Therefore, a test of SEP constrains the direction of a modified theory of gravity.

The violation of SEP is measured with the Nordtvedt parameter $\eta$, which is $\eta=4 \beta -\gamma -3$ using the parametrized post-Newtonian (PPN) parameters, $\beta$ and $\gamma$ (if GR holds, $\eta=0$) \footnote{Parametrized post-Newtonian (PPN) formulation is the standard expression of post-Newton limit of metric theories of gravity, {\it i.e.}  with weak gravity limit of any theories.  There are 10 parameters in the full PPN formulation, but in the primitive version (space-time around a static, spherical, non-rotating mass) space-time metric is described as 
\begin{eqnarray}
g_{00} &=& -1+2U -2\beta U^2, \label{postNewtong00}\\ g_{0j}&=& 0, \\  g_{ij}&=& (1+2\gamma U) \delta_{ij} 
\end{eqnarray}
where $U=M/r$ where $M$ is the source mass and $r$ is the distance from the source.  In GR, $\beta=\gamma=1$. }.

Since testing SEP requires self-gravitating body, possible experiments require bigger masses or long baselines, therefore 
experiments are mostly done using spacecraft or by observing astronomical phenomena. 
\begin{itemize}
\item Lunar laser ranging, which tracks the Moon orbit including gravity of Sun by 
measuring the Earth--Moon distance to high precision, provides a variety tests of gravity; such as WEP, SEP, constancy of gravitational constant $G$, the inverse square law, geodesic precession, gravito-magnetism, and so on. 
The current measurement with a millimeter precision shows $\eta = (4.4 \pm 4.5) \times 10^{-4}$. 
(Refs. \cite{Williams2004, Williams2009, Murphy2012})
\item Several tests of SEP have been proposed, such as 
a tracking BepiColombo, the spacecraft for Mercury \cite{Milani2002},
an observation of a triple pulsar \cite{Shao2016},  
measurement of spacecraft ranging at the Sun-Earth Lagrangian points \cite{Congedo2016}. 
\end{itemize} 

%--------------------------------------------------------------------------------------
\subsection{Gravitational Redshift}\label{sec_gravredshift}
%--------------------------------------------------------------------------------------
When light comes out against the gravity of a star, its frequency decreases, which is called {\it gravitational redshift}.
%This can also be interpreted as a phenomenon in which light resists gravity and loses energy. 
Changes in frequency can be regarded as changes in time scale due to the gravitational field.

\begin{figure}[thpb]
\centerline{\psfig{file=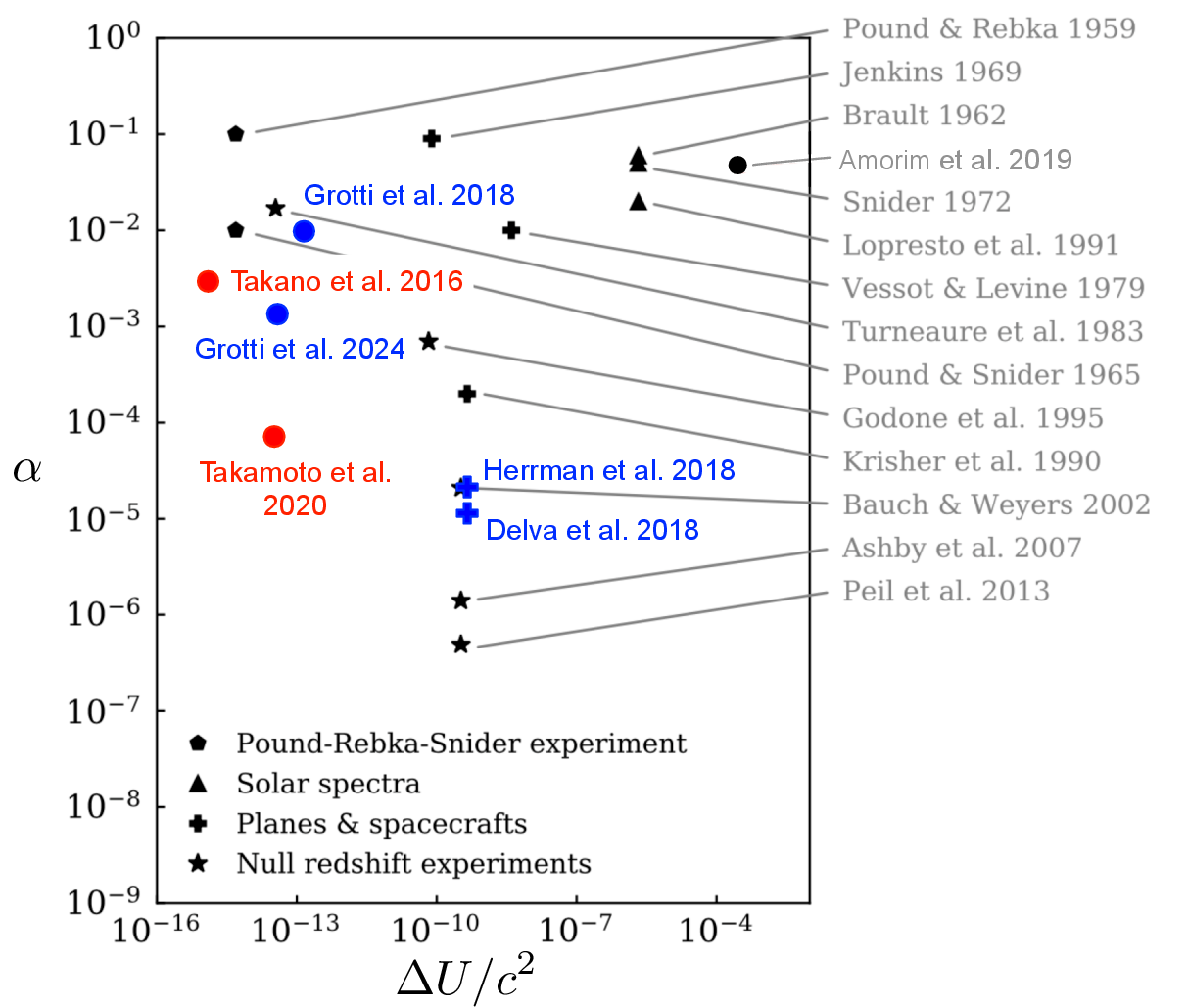,width=10cm}}
\vspace*{8pt}
\caption{Comparison of selected tests of the LPI with gravitational redshift in the plane of the variation of potential, $\Delta U/c^2$, and the measured limit on $\alpha$. 
We update Fig. 3 in Ref. \cite{Amorim2019}. 
The different symbols mark the Pound-Rebka-Snider experiments \cite{PoundRebka1959,PoundSnider1965}, 
tests from solar spectral lines \cite{Brault1962,Snider1972,LoPresto1991}, 
tests on rockets and spacecrafts \cite{Vessot1980,Smarr1983,Delva2018,Herrmann2018,Hafele1972,Krisher1990} , 
atomic clock experiments \cite{Turneaure1983,Godone1995,Bauch2002,Ashby2007,Peil2013}, and optical lattice clocks\cite{Takano2016,Takamoto2020,Grotti2018,Grotti2024}. 
Colored points are of the largest uncertanites in $\alpha$, while black ones are from Ref.  \cite{Amorim2019}.
 \label{fig_gravredshift}}
\end{figure}

According to EEP, there is no distinction between experiments in a gravitational field and experiments in uniformly accelerated motion. Therefore, verifying the gravitational redshift is equivalent to comparing experiments in different inertial coordinate systems, and is a test of LPI. 
\begin{itemize}
\item 
The first redshift measurement was carried out in a series of
Pound-Rebka-Snider experiments \cite{PoundRebka1959,PoundSnider1965} in the early 1960s. 
They measured the gravitational M\"ossbauer effect using gamma rays emitted from ${}^{57}$Co at an altitude difference $\Delta h =22.5$~m, and reported 
$| \alpha | < O (10^{-2}) $. 
\item Many attempts were made for measuring gravitational redshift from the Sun, but it was not so simple since solar spectral lines include the effect of photosphere and chromosphere.  By means of special absorption lines, the measurements were reported by Brault\cite{Brault1962}, Snider\cite{Snider1972} and by LoPresto et al\cite{LoPresto1991}.
\item 
The Gravity Probe A mission \cite{Vessot1980,Smarr1983} obtained $| \alpha | \approx 1.4 \times 10^{-4} $ using a hydrogen maser in a spacecraft launched to 
$\Delta h =10,000$~km. 
\item 
Using two Galileo satellites that accidentally took elliptic
orbits with a height difference of $\Delta h \approx 8,500$~km, new constraints were
reported  as $\alpha = (0.19 \pm  2.48) \times 10^{-5}$ (ref. \cite{Delva2018}) and $\alpha = (4.5 \pm 3.1) \times 10^{-5}$ (ref. \cite{Herrmann2018}). 
\item Comparing a transportable clock in the middle of a mountain and a laboratory clock, with $\Delta h \approx 1,000$~m (ref. \cite{Grotti2018}), Grotti  {\it et al.} \cite{Grotti2018} obtained  $\alpha \approx 10^{-2}$. Grotti {\it et al} also reported their ground level test (ref. \cite{Grotti2024}), which can be estimated as $\alpha \approx 10^{-3}$.

\end{itemize}
There have been many other reports of measurements, such as 
atomic clocks mounted on a civilian aircraft \cite{Hafele1972},  
monitoring of the ultrastable crystal oscillator in Voyager 1 spacecraft during its encounter with Saturn \cite{Krisher1990}, 
the triplet infrared spectrum of oxygen atoms on the surface of the sun \cite{LoPresto1991}, 
comparison of atomic clocks \cite{Turneaure1983,Godone1995,Bauch2002,Ashby2007,Peil2013}, 
the star orbit around the super-massive black-hole of the Milky Way galaxy\cite{Amorim2019}. 
These obtained limits are summarized in Fig.~\ref{fig_gravredshift}.

The uncertainty of $\alpha $ is mainly given by $(c^2/\Delta U) (\delta \nu /\nu_1) $,  suggesting 
that accurate frequency measurement of clocks ($u_c = \delta \nu /\nu_1$ ) is at the  
heart of the endeavor, in particular, for ground experiments with $\delta h$ less than a kilometer, as $\Delta U$ is nearly four orders of magnitude smaller than the space experiments.  
As we explain in \S 4, our experiments using OLC are: 
\begin{itemize}
\item 
A comparison of OLCs  at RIKEN and The University of Tokyo \cite{Takano2016} with $\Delta h =15$~m has so far demonstrated 
$\alpha = (2.9 \pm 3.6) \times  10^{-3}$, limited by $u_c = 5.7 \times 10^{-18}$. 
%Constraining $\alpha $ to better than $10^{-3}$ on the ground has remained uninvestigated, as it requires outstanding clock accuracy or height differences.
\item A comparison of OLCs at Tokyo Skytree \cite{Takamoto2020} with $\Delta h = 450$~m shows $\alpha = (1.4\pm 9.1) \times  10^{-5}$. 
\end{itemize}
These two results are marked with red points in Fig. \ref{fig_gravredshift}.

%=================================================
\section{The experiments with optical lattice clocks \label{sec:SkyTree}}
In this section, two experiments are presented that demonstrate frequency comparisons between OLCs with long distance (\S \ref{subsec_RIKENUT}) and large height differences (\S \ref{subsec_Skytree}) \cite{Takamoto2020}, which are the experiments of gravitational redshift (\S \ref{sec_gravredshift}), a part of testing the equivalence principle.
Such comparison of OLCs is a key technology for future applications in chronometric leveling.

\subsection{The experiment between RIKEN and Univ. of Tokyo}\label{subsec_RIKENUT}
According to the gravitational time dilation effect predicted by GR, a clock at a lower elevation ticks slower than a clock at a higher elevation. 
A comparison of two clocks' frequencies with an uncertainty of $10^{-18}$ reveals a height difference of the clocks by 1 cm on the ground. 
Such clock-based height measurements are called ``chronometric leveling'' \cite{Vermeer1983}. 
A network of clocks \cite{Takano2016,Schioppo2022} would serve as benchmarks, allowing high-resolution geopotential mapping and monitoring of gravitational potential to detect crustal deformation \cite{Tanaka2021}.
\begin{figure}[tbp]
%\vspace*{-2.1cm}
\begin{center}
\hspace*{0.6cm}
\includegraphics[width=0.8\linewidth]{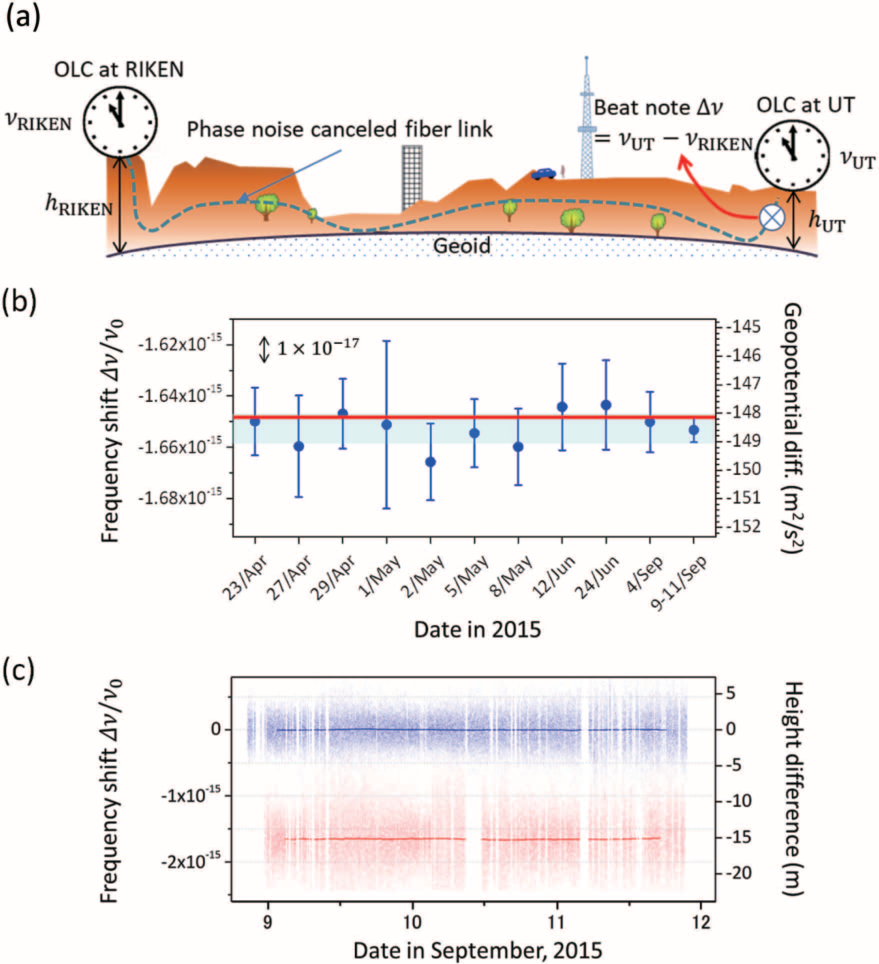}
\end{center}
%\vspace*{2cm}
\caption{Chronometric leveling demonstrated by frequency comparisons between RIKEN and UT. 
The frequency difference of the remote clocks $\Delta \nu=\nu_{\rm UT}-\nu_{\rm RIKEN}$ measures the height difference $\Delta h = h_{\rm UT}-h_{\rm RIKEN}$ between the clocks as $\Delta \nu / \nu_{\rm UT} =(U_{\rm UT}-U_{\rm RIKEN})/c^2 \simeq g \Delta h/c^2 = 1.1\times 10^{-18}$  $\Delta h/\mathrm{cm}$, where $U_i$ and $h_i$ are the geopotential and height for the clock at $i =$ UT and RIKEN, respectively.
\label{fig_RIKENUT}}
\end{figure}

To demonstrate such chronometric leveling, we performed a remote frequency comparison of OLCs developed at RIKEN and the University of Tokyo (UT) \cite{Takano2016}, as illustrated in Fig.~\ref{fig_RIKENUT}(a).
The clocks were operated at each site, and their clock frequencies were compared using an optical fiber link.
The 15-km distance between the sites was connected by an optical fiber with a length of 30 km.
The clock frequency was transferred from RIKEN to UT by a phase-noise canceled optical fiber link, and the frequency difference between the clocks $\Delta \nu = \nu_{\rm UT} - \nu_{\rm RIKEN}$ was measured at UT.

With 11 measurements over half a year (Fig.~\ref{fig_RIKENUT}(b)), the fractional frequency difference between the clocks was determined to be $\Delta \nu/\nu_{\rm UT} = (-1652.9 \pm 5.9) \times 10^{-18}$.
The frequency offset measures the gravitational redshift for a height difference between the sites ($\Delta h = h_{\rm UT} - h_{\rm RIKEN} \sim -15~\mathrm{m}$).
The result was consistent with the height difference obtained by the conventional leveling results (red line in Fig.~\ref{fig_RIKENUT}(b)) within the uncertainty, confirming the validity of relativistic geodesy with clock comparisons.
From the obtained results, the parameter $\alpha$ (see eq.~(\ref{eq_alpha})) was derived to be $\alpha = (2.9 \pm 3.6)\times 10^{-3}$.
%, which validates GR by three orders of magnitude.
Such remote clock comparison, unlike conventional leveling surveys, allows real-time monitoring of the gravitational potential (Fig.~\ref{fig_RIKENUT}(c)) and will become an important geodetic technique in the future.
%If clocks are networked by optical fiber networks, height references can also be constructed, and optical fiber links are being built in Europe for this purpose.

\subsection{The experiment at Tokyo Skytree}\label{subsec_Skytree}
Next, we describe the experiment at Tokyo Skytree in 2018 \cite{Takamoto2020}, by setting two transportable clocks with the height difference of $\Delta h \approx 450$~m. 
%The use of transportable optical lattice clocks (\S \ref{subsec_TOC}) allows clock comparisons at a location dedicated to the measurement purpose.
For the purpose of testing GR, it is advantageous to choose a location with a large height difference and in the same location, where the clocks can be easily linked and the surveying can be done more accurately.
Therefore, we performed an experiment at Tokyo Skytree, where a height difference of 450~m is accessible by a 700~m fiber link, a suitable location for the test of GR on the ground.

\begin{figure}[tb]
%\vspace*{-2.1cm}
\begin{center}
\hspace*{0.6cm}
\includegraphics[width=\linewidth]{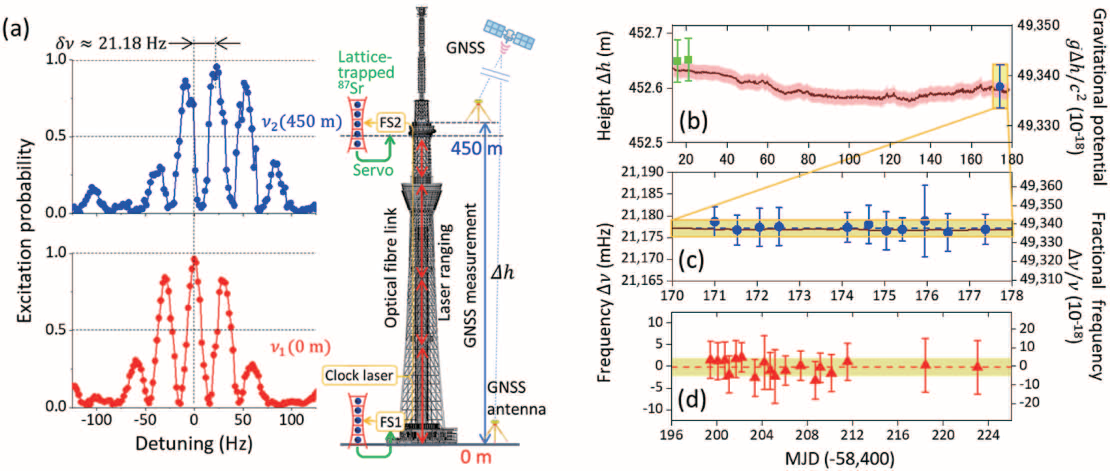}
\end{center}
%\vspace*{2cm}
\caption{Test of GR at Tokyo skytree. 
(a) The gravitational potential difference is investigated by two different methods of laser ranging and GNSS leveling complemented by spirit leveling and a gravimeter. 
Ramsey spectra measured in the observatory floor (blue circles) and the ground floor (red circles) give the gravitational redshift of $\sim 21.18$~Hz.
(b) Height difference $\Delta h$ between the clocks measured by GNSS (green squares), laser ranging (brown line) and the beat note of the clocks (blue circle) after applying systematic corrections. 
The red-shaded region shows the uncertainties of height measurements by laser ranging. 
(c) The gravitational redshift measured at Skytree (blue circles) is consistent with the gravitational
potential difference measured by the laser ranging and gravity measurements (brown line). 
(d) The beat note of two clocks set at the same height (red triangles) demonstrates an excellent reproducibility  $(-0.3 \pm 4.7) \times 10^{-18}$ of the clocks. 
MJD denotes the modified Julian date.
\label{fig_Skytree}}
\end{figure}

The experimental setup is shown in Fig.~\ref{fig_Skytree}(a).
Two transportable OLCs were set at Tokyo skytree; one on the ground floor and the other on the observatory floor. 
The clock laser was set on the ground floor, and the signal was sent to the observatory floor via a phase-noise-canceled optical fiber to interrogate the clock transition. 
The left panels of Fig.~\ref{fig_Skytree}(a) show the Ramsey spectra measured at the ground floor and at the observatory floor, respectively, with a pulse duration of 11~ms and a free evolution time of 20~ms. 
The frequency shift of $\Delta \nu =\nu_2 - \nu_1 \approx 21.18$~Hz corresponds to a gravitational redshift of the clock frequency ($\nu_1 \approx 429.228$~THz) for a height difference of $\sim 450$~m. 
The clock laser frequencies $\nu_1$ and $\nu_2$ are stabilized to the peak of the central fringe of respective Ramsey spectra with a free-evolution time of 40~ms using frequency shifters. 
While running the clocks, the geopotential difference between the clocks was also investigated using the GNSS leveling and laser ranging complemented by spirit leveling and gravity measurements. 

Figure \ref{fig_Skytree}(b) summarizes the geopotential measurements performed by the clocks, GNSS leveling, laser ranging, and gravimeter. 
We conducted GNSS leveling for five days in October 2018 to obtain $\Delta h= (452.650 \pm 0.039)$~m. 
This agrees with the simultaneous laser ranging value of $\Delta h=(452.631 \pm 0.013)$~m within $1\sigma$ uncertainty, validating the consistency of the height measurements. 
The laser ranging continuously monitors the long-term height variation of the tower corresponding to a temperature variation of 10~${}^\circ$C over 6 months.
Figure \ref{fig_Skytree}(c) presents 11 measurements taken to determine the gravitational redshift for $\Delta h$.
After the measurements at Skytree, we transported the system back to RIKEN and compared the two clocks at the same height (Fig. \ref{fig_Skytree}(d)). 
The measured fractional beat note confirms the reproducibility of the clocks. 
During the measurement period, %`modified Julian date' (MJD) 58571--58577, the gravitational redshift is observed to be $\Delta \nu / \nu=(49,337.8 \pm 4.3)\times 10^{-18}$, while 
the gravitational potential difference measured by laser ranging gives $\overline{g} \Delta h / c^2 = (49,337.1 \pm 1.4) \times 10^{-18}$ with height difference $\Delta h=(452.596 \pm 0.013)$~m and gravitational acceleration $ \overline{g} \sim (9.797248 \pm 0.000024)$~m~s$^{-2}$. 
These results indicate a value of $\alpha= (1.4 \pm 9.1)\times 10^{-5}$, giving the best constraint on the gravitational redshift on the ground. 
This result is comparable to a space-borne experiment using atomic clocks on satellites \cite{Delva2018,Herrmann2018}
and complementary as it covers the short range (450~m from the surface) in addition to the already covered long range ($10^4$ km) \cite{Delva2018,Herrmann2018} for LPI tests at the $10^{-5}$ uncertainty level.
A further constraint on $\alpha$ can be set by improving the clock's uncertainty.

%\clearpage
%=================================================
\section{Future possible applications \label{sec:Future}}
%=================================================
Many authors have proposed applications of OLCs. See e.g., a review by  Ludlow {\it et al.} \cite{Ludlow2015}. 
In this section, we provide an outlook on implementational, technological and scientific viewpoints, and introduce our proposal of new gravitational wave detector using OLCs.

%--------------------------------------------------------------------------------------
\subsection{Outlook}
%--------------------------------------------------------------------------------------
\paragraph{Implementational viewpoints}~\\
The unique feature of OLCs compared to conventional atomic clocks lies in their high stability, enabling real-time monitoring of dynamic changes of gravitational redshift.
As the stability of OLCs relies on the highly stable clock lasers that require half a meter long or cryogenic reference cavities, pursuing high stability is not compatible with the transportability of the OLCs.
To achieve high clock stability less relying on the stability of the clock laser, we proposed a ``longitudinal spectroscopy" \cite{Katori2021}, which allows continuous interrogation of the clock transition.
The main challenge to experimentally constitute the continuous clock is realizing the continuous source of ultracold atoms. We developed a continuous outcoupling of atoms into a moving lattice\cite{Takeuchi2023}, where atoms were further guided to the orthogonal direction to be free from the spontaneously emitted photons during laser cooling of atoms\cite{Okaba2024}.

We have also started to explore real-world applications that take full advantage of the clocks' accuracy and stability.
To this end, it is crucial to develop technologies to downsize the clock and improve both transportability and robustness.
We are working jointly with industrial partners to develop a compact clock system for future commercialization.
A prototype system was developed with a total volume of 250 L, a quarter of our previous clock system.
The volume of the system is already close to being placed in space.

\paragraph{Redefinition of the second}~\\
One near-future issue is a redefinition of the second.  Since 1967, the International System of Units (SI), the second, is defined by Cs atomic clocks with an uncertainty of $\sim 5 \times 10^{-16}$.
Both International Atomic Time (TAI) and Coordinated Universal Time (UTC) are currently maintained with a large ensemble of Cs atomic clocks and hydrogen masers, of which instability is about $4\times 10^{-16} $ over 30 to 40 days before the calibration.  
Optical clocks will improve this situation by two orders of magnitude. 
Redefinition of the second by OLCs is lively discussed \cite{Riehle2018}.

\paragraph{Chronometric leveling}~\\
As was shown in \S 4, OLCs can identify a tiny gravitational redshift due to geopotential (potential of gravitational plus centrifugal components). 
Therefore, OLCs can also be a precise tool for measuring geopotentials.
% Such clock-based geopotential (height) measurements are called “chronometric leveling” \cite{Vermeer1983}. 
%A network of clocks7-8 would serve as benchmarks, allowing high-resolution geoid mapping and monitoring of gravitational potential to detect crustal deformation.6
High accuracy of OLCs provides a height reference by connecting clocks worldwide.
Moreover, the unique feature of OLCs compared to conventional atomic clocks is their high stability, enabling real-time monitoring of dynamic changes of gravitational redshift.
Real-time geopotential measurements at the centimeter level will open up new applications in future geodetic technology, such as improvement of geodetic leveling \cite{Denker2017}, including seismology and volcanology \cite{Tanaka2021}.
In the future, we expect compact transportable OLCs to be deployed as a social infrastructure and networked by optical fibers, to serve for geophysics applications, such as observing crustal deformation caused by plate movement and volcanic activity.

%\newpage
%--------------------------------------------------------------------------------------
\subsection{Test of PPN potential at the second order}\label{subsecISS}
%--------------------------------------------------------------------------------------

We next consider the possibility of testing the PPN  potential at the second order. 
Since the second-order effect can be observed more clearly between separated positions, we consider to locate one OLC in space. 
From the expansion of eq. (\ref{postNewtong00}), the clock in space (distance $r_A$ from the center of the Earth) ticks faster than that on the surface of the Earth (distance $r_B$) with the ratio $\sqrt{1-2U_A + 2\beta U_A^2}/\sqrt{1-2U_B + 2\beta U_B^2}$, where $\beta=1$ for GR. The latest test indicates $|\beta -1 | = (0.2 \pm 2.5) \times 10^{-5}$ by the perihelion shift of Mercury \cite{Will1993Book}.

If we locate one clock at a Kepler circular orbit around the Earth, then the difference of the time-ticks between the one in space and the one on the Earth surface (at Tokyo,  $35^\circ$ N) is as shown in Table \ref{table2}. The numbers include both special and general relativistic effects, but do not include the shift of orbital radius for the International Space Station (ISS). (ISS orbits around the Earth with the height between 250 and 400~km from the surface of the Earth.  Due to the friction, ISS loses its height $\sim$ 1~km a day, and shorten its circulation period  $\sim$ a few seconds a day.)

Table \ref{table2} indicates that the second-order effect is 10th-order smaller than that of the first-order. 
The direct comparison of elapsed time at the orbit of GPS satellite with that on the surface of the Earth, $\Delta t/t \sim 4.5 \times 10^{-19}$, is critically reachable with current technology, while we have to separate the second-order effect from the background. 
If we make the orbit at $\sim 3178$~km from the surface, then the total time dilation due to special relativity and the first order PPN potential offset, and the separation problem of the second order effect would be tractable. (But the level $\Delta t/t \sim 2.7 \times 10^{-20}$ is required.)

Since most higher-order curvature
modified gravitational theories ($f(R,\phi)$ theories, which try to solve the inflationary Universe and/or the accelerating expansion of the Universe) show the difference with GR at the second-order PPN potential level, the constraining $\beta$ is interesting as a test of general relativity.  The measure of $\beta$ will also be a part of test of SEP (See \S 2).

\begin{table}[htb]
\tbl{Special and general relativistic time dilation between the one in height and the one on the surface. The signature $+$ means delay from the surface of the Earth. We assume Kepler circular motion in space, and the surface clock is supposed at Tokyo, Japan (N 35 degree).  SR effect is due to the satellite velocity and the rotation of the Earth surface. The columns PPN 1st/2nd are the contribution of PPN potential terms of first and second order (with $\beta = 1$ in (\ref{postNewtong00})), respectively. The total difference at the GPS satellite is different signature than the other cases, which is due to the high speed of GPS satellite. }
{\begin{tabular}{@{}lcccc@{}} %\toprule
\hline 
orbit 
& $\Delta t/t $ (total, SR+GR) 
& $\Delta t/t $ (SR) 
& $\Delta t/t $ (PPN 1st) 
& $\Delta t/t $ (PPN 2nd)
\\
\hline
GPS (@20184km) 
& $-4.4569\times 10^{-10}$ 
& $+8.2676\times 10^{-11}$ 
& $-5.2836\times 10^{-10}$ 
& $-4.5560\times 10^{-19}$
\\
@3177.963km
&  0
&  $+2.3124\times 10^{-10}$ 
&  $-2.3124\times 10^{-10}$ 
&  $-2.6810\times 10^{-20}$
\\
ISS (@400km) 
&  $+2.8531\times 10^{-10}$ 
&  $+3.2634\times 10^{-10}$ 
&  $-4.1034\times 10^{-11}$ 
& $-5.5380\times 10^{-20}$
\\
%ISS (@300km) 
%& $+3.0000\times 10^{-10}$ 
%& $+3.3124\times 10^{-10}$ 
%& $-3.1236\times 10^{-11}$ 
%& $-4.2463\times 10^{-20}$
%\\
%Tokyo Skytree(@450m) 
%& $+3.4678\times 10^{-10}$ 
%& $+3.4683\times 10^{-10}$ 
%& $-4.9055\times 10^{-14}$ 
%& $-6.8242\times 10^{-23}$
%\\
\hline
\end{tabular}
\label{table2}}
\end{table}

%--------------------------------------------------------------------------------------
\subsection{Gravitational Wave Observatory at Space: Proposal of INO}\label{subsecINO}
%--------------------------------------------------------------------------------------
In this subsection, we introduce our proposal of new gravitational wave (GW) observatory\cite{Ebisuzaki2019}. 

The observation of GW has opened new era for physics and astronomy. 
The first detection was in 2015 (GW150914)\cite{GW150914PRL}, whose source was a coalesce of a binary black-holes (BHs). 
The first detection of GW from a binary neutron stars (GW170817)\cite{GW170817PRL} was followed up by many other 
observatories (with the electromagnetic waves from radio to gamma-ray and by neutrino observation), 
which was the starting moment of ``multi-messenger astronomy". 
So far, LIGO-Virgo-KAGRA collaboration has announced 90 detections of GW \cite{GWTC3}, and we can discuss not only 
the orbital parameters of the sources but also several astrophysical aspects such as 
the details of high-energy events, the equation of state of nuclear matter, cosmology, and the validity of gravitational theories.

Among the unsolved problems in the Universe, however, the growth process of large BHs is left untouched. 
Almost all of the galaxies in the Universe have super-massive black holes (SMBHs) in their center, whose mass is over $10^6 M_\odot$, but we do not know how to form such BHs. 
One of the plausible scenario is the hierarchical growth model of the stars in the galaxy, that is, a SMBH was formed by mergers of small BHs.   If this is a route, the mergers of intermediate-mass black holes (IMBHs) are expected to be the sources of GW\cite{Matsubayashi:2004it, Shinkai:2017gb}, which are not observed by the current ground-based GW detectors since such mergers produce low-frequency GWs below 1~Hz those are in the range of seismic vibration of the Earth.

There are several projects of placing GW detectors in space (see a review by Ni \cite{WeiTouNi2016, WeiTouNi2024}). 
The project of ``Laser Interferometer Space Antenna" (LISA) by ESA \cite{eLISA} is the plan of constructing laser interferometer in space with the arm length $1.0 \times 10^6$ km, targeting mainly at milli-Hz range of GW. 
Locating three spacecrafts at Earth-like solar orbits with $10$-degree lag
with drag-free flight motion, and using the light-transponder technique, ESA plans to realize the system in the middle of 2030s.  
Japanese group proposed ``DECi-hertz Interferometer GW Observatory" (DECIGO/B-DECIGO) project \cite{Nakamura:2016cn}, which plans to construct Fabry-Perot laser interferometer with 1000~km (100~km) arm length, with three spacecrafts on the Sun-Earth orbit (around the Earth orbit) with drag-free flight motion.  Their main target is deci-Hz range of GW. 

\begin{figure}[thb]
\begin{center}
\includegraphics[width=0.75\linewidth]{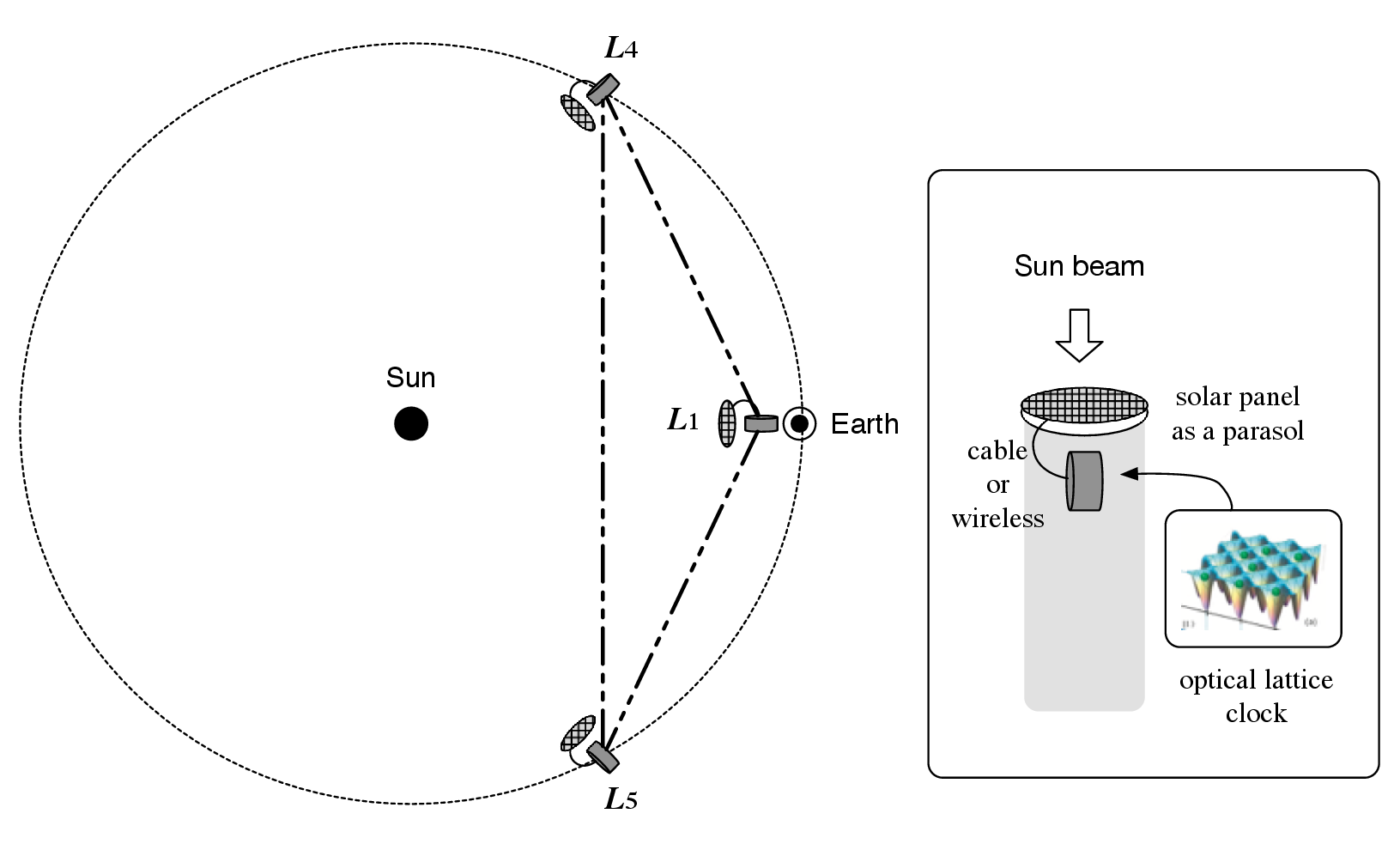}
\end{center}
\caption{A planned location of the INO spacecrafts: Lagrangian points $L_1$, $L_4$ and $L_5$ of the Sun-Earth orbit. The $L_1$ is at 
1/100 A.U. from the Earth, while $L_4$ and $L_5$ forms equilateral triangle with the Sun and the Earth respectively; the distance between $L_1$--$L_4$($L_5$) is 1 A.U., while that of $L_4$--$L_5$ is $\sqrt{3}$ A.U.
Two-frequency radio or light will be used for communication between spacecrafts.  The inset explains that the solar panel of the  spacecrafts is separated as a parasol from the main body, in order to prevent acceleration noise due to solar wind. 
[Fig.1 of Ebisuzaki {\it et al.} \cite{Ebisuzaki2019}]
\label{fig_INO_fig1}}
\end{figure}

Space-borne interferometers such as LISA or B-DECIGO require significant technical breakthroughs.  
We\cite{Ebisuzaki2019} propose an alternative method for detecting low-frequency GWs, technically feasible with the current technologies.  Our idea is to locate three spacecrafts at A.U. scales (say at L1, L4 and L5 of the Sun-Earth orbit), which load 
OLCs (see Figure \ref{fig_INO_fig1}). 
By comparing the time each other, applying the principle of the Doppler tracking, we can detect the passage of GWs of mHz range. We named this proposal {\it ``Interplanetary Network of Optical Lattice Clocks"}, with the abbreviation {\it INO}. 
The acronym, INO, is named after Tadataka Ino (1745-1818), a Japanese astronomer, cartographer, and geodesist, who made precise map of Japan two centuries ago\footnote{Ino spent 16 years surveying the coastline of Japan while making astronomical observations, and produced the precise map of Japan. His motivation for creating the map was to determine the exact length of one degree of latitude and thus the size of the Earth.}. 

%The ideas of locating spacecrafts at A.U. scales have already been appeared \cite{AMIGO1,SuperASTROD}.  For example, Ni introduced a mission concept of ``Astrodynamical Space Test of Relativity using Optical Devices" (ASTROD) \cite{1212.2816,AMIGO1} and Super-ASTROD\cite{SuperASTROD}.  ASTROD is to locate spacecrafts at Sun-Earth L3/L4/L5, while Super-ASTROD is to locate spacecrafts at Sun-Jupiter L3/L4/L5 together with at Sun-Earth L1/L2, and to probe primordial GWs at $10^{-6}$--$10^{-3}$ Hz by Doppler tracking using laser pulse ranging and precision optical clocks \cite{WeiTouNi2016}.

%The ideas of using atomic clocks for detecting GW have also already been appeared \cite{Tinto2009, 1501.00996, Vutha2015, Kolkowitz:2016gx,SuWangWangJetzer}.   For example, Su et al. \cite{SuWangWangJetzer} proposed a similar idea naming ``Double Optical Clocks in Space" (DOCS). They proposed to place two optical lattice clocks at the Lagrange points of the Earth-Moon orbit, and link them with the Earth by radio. 

%---------------  figure 1 --------------->>>>
%---------------  figure 1 --------------<<<<

%---------------  figure 2 --------------->>>>
\begin{figure}[bh]
\begin{center}
\includegraphics[width=0.75\linewidth]{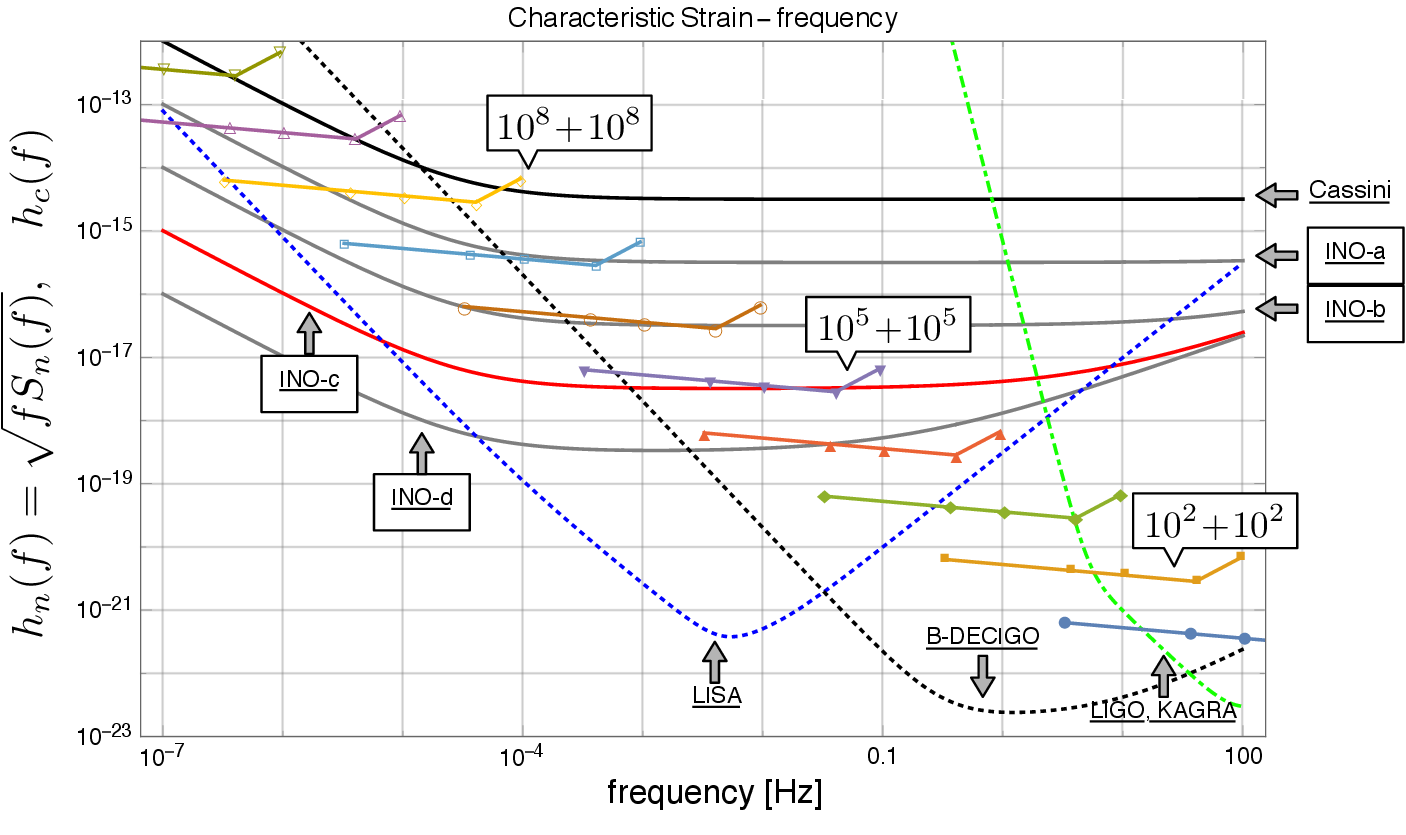}
\end{center}
\caption{
Sensitivity of Doppler-tracking spacecrafts and expected strains of GW.
The most upper solid curves indicates the sensitivity of Cassini spacecraft (2001), while the other solid curves are 
those of 1$\sim$4-order improved version (we named INO-a, INO-b, INO-c and INO-d, respectively). 
The dotted line is the sensitivity curve of LISA. Almost horizontal lines with symbols indicate the characteristic strain of GW from a merger of equal-mass binary BHs at 1~Gpc. 
Each line is for the inspiral phase; starts from its separation 50 times of the event horizon radius, and ends at their merger  (frequency moves up higher for smaller separation).  [Fig.2 of Ebisuzaki {\it et al.} \cite{Ebisuzaki2019}]
\label{fig_INO_fig2}}
\end{figure}
%---------------  figure 2 ---------------<<<<

In the reference \cite{Ebisuzaki2019}, we discussed feasibilities and technological new idea together with detectable distance of the detectors and GW sources counts.

We estimate the reachable sensitivity for GW detection with current known technologies.  
In order to make the most feasible discussion, we do not consider to use drag-free control, nor precise laser control, but simply apply the advanced OLC to the Doppler-tracking method. 
The sensitivity of the Doppler-tracking method is well understood by the report of Cassini spacecraft 
\cite{Armstrong:2006hw,Armano:2017ij}, which keeps the best record as 
$h_{\rm n}\sim3\times10^{-15}$ at 10$^{-4}$~Hz, where 
$h_{\rm n}$ is the noise amplitude, which is given by the square root of the combination of 
the power spectrum of the noise times frequency $f$. % \cite{Berry:2015wk}. 
The noise amplitude is the standard quantity since it can be compared directly with the 
characteristic strain $h_{\rm c}$ which expresses the strength of the GW signal. 
Cassini's sensitivity showed the curve $f^{-1}$ below 10$^{-4}$~Hz.  
%This behavior is in contrast with $f^{-2}$ which is seen in the drag-free spacecraft.

The origins of noise in Cassini are identified mostly from the uncertainty of the atomic clock and from the fluctuation of troposphere of the Earth \cite{Armstrong:2006hw}. 
If we use the OLC instead of the Cs atomic clock, and let the spacecrafts communicate
each other directly, and with a Sun-beam shield, 
the sensitivity will be dramatically improved: 
%From the Table.4 in \cite{Armstrong:2006hw}, we estimate that 
the three or four-order improved version of Cassini spacecraft ({\it i.e.} the minimum sensitivity is around 
$h_{\rm n}=10^{-17}$ or $10^{-18}$) will be available.

Figure \ref{fig_INO_fig2} shows the sensitivity curve of Cassini spacecraft and their one to four-order improved version (we named INO-a, INO-b, INO-c and INO-d, respectively), together with that of LISA, B-DECIGO and advanced LIGO/KAGRA.  
Since the frequency dependence at the lower frequency is different, INO achieves the same sensitivity with LISA at 10$^{-5}$~Hz, and better than LISA in the range less than that.

In Fig. \ref{fig_INO_fig2}, we also plot the characteristic strain of the GW ($h_{\rm c}$) from a merger of the binary BHs with its distance 1~Gpc from the Earth. 
We plotted for mergers of equal-mass BHs for several different masses.  Each line starts from its frequency when the binary's separation is 50 times of their event horizon radius, and ends at the frequency when they merge. 

We see that the mergers of SMBHs of 10$^7\sim$10$^8 M_{\odot}$ produce GW around 10$^{-4}$~Hz, which is detectable with INO at the signal-to-noise ratio 10. 
Our proposal does not reach  the best sensitivities and detectable distance than so-far proposed concepts.  However, we shall show that our concept is enough for testing a SMBH-formation scenario with the currently available technologies within a certain operation period.  See Ebisuzaki {\it et al.} \cite{Ebisuzaki2019} for more detail. 

%We \cite{Ebisuzaki2019} have proposed to place OLCs in space and to use Doppler tracking method for detecting low-frequency gravitational wave below 1 Hz.   Our idea is to locate three spacecrafts at one A.U. distance (say at L1, L4 \& L5 of the Sun-Earth orbit), and apply the Doppler tracking method by communicating ``the time" each other.  Applying the current available technologies, we obtain the sensitivity for gravitational wave with three or four-order improvement ($h_{\rm n}\sim 10^{-17}$ or $10^{-18}$ level in $10^{-5}$Hz -- $1$ Hz) than that of Cassini spacecraft in 2001.  This sensitivity enables us to observe black-hole mergers of their mass greater than $10^5 M_\odot$ in the cosmological scale.  Based on the hierarchical growth model of black-holes in galaxies, we estimate the event rate of detection will be 20-50 a year. 

%=================================================
\section{Summary \label{sec:Outlook}}
%=================================================
The development of the optical lattice clocks (OLCs) has enabled precise measurement of gravity.  As was demonstrated by our experiments at RIKEN-U.Tokyo, and at Tokyo Skytree Tower, we can measure the height difference with an uncertainty of 5 cm using fiber-linked two clocks.  This precision will be improved by an order of magnitude in the near future.

From the viewpoint of theoretical physics, such implementations mean that the various tests of physics can be realized.  
As an example, we outlined an idea to use OLCs in space for testing second-order effects of PPN potential to time-ticks which enables to further test of strong equivalence principle, and for detecting low-frequency gravitational waves which targets the problem how super-massive black-holes were formed. 

On the other hand, from the technological viewpoint, OLC opens the field of ``relativistic" geodesy. 
The precise determination of gravitational potential will be used in seismological and volcanological studies \cite{Tanaka2021}.  
%By monitoring geopotential changes by OLCs, an uncertainty of atmospheric delay models used in GNSS analyses can be investigated, and the effects of tides and surface loads can be revealed. 

Einstein mentioned that the theories of relativity will not contribute to improving our lives.  However, the development of OLC is now close to changing our fundamental method of measuring ``height" and ``velocity" based on the relativity theories.  %Therefore Einstein's words would be denoted as one of his pessimistic thoughts or as his additional mistake by future historians.   
The fast development in science and technology overreached Einstein's expectation.

%--------------------------------------------------------------------------------------
\section*{Acknowledgment}
%--------------------------------------------------------------------------------------

This work received support from the Japan Society for the Promotion of Science (JSPS) Grant-in-Aid (grant No. 24K07029, HS) and the Japan Science and Technology Agency (JST) Mirai Program Grant No. JPMJMI18A1.

%This work received support from a Japan Society for the Promotion of Science (JSPS) Grant-in-Aid for Specially Promoted Research (grant no. JP16H06284) and Japan Science and Technology Agency (JST)-Mirai Program grant no. JPMJMI18A1. H.S. acknowledges support from JSPS KAKENHI grant no. JP17H06358. We thank Shimadzu Corporation for development of control electronics for the laser system, Geospatial Information Authority of Japan for GNSS, levelling and gravity measurements, Tobu Tower Skytree Co. for support of the experiments, J. Fort\UTF{00E1}gh and L. S\UTF{00E1}rk\UTF{00E1}ny for the loan of wavelength meters, Y. Takahashi from Citizen Watch Co. for development of a laser system, M. Kokubun for support with electronics, K. Araki for designing control electronics, T. Takahashi, H. Ichikawa and A. Gomyo for laser ranging measurements and A. Hinton for reading the manuscript.

\end{document}